\documentclass[twocolumn,dvipsnames]{aastex62}
\pdfoutput=1 
\usepackage{amsmath,amstext}
\usepackage{apjfonts}
\usepackage[figure,figure*]{hypcap}
\usepackage{ulem}
\usepackage{xspace}

\usepackage{graphicx}
\usepackage{natbib}
\usepackage{amssymb}
\usepackage{amsmath}
\usepackage{array,booktabs}
\usepackage{mathptmx}
\usepackage{booktabs}
\usepackage{hyperref}
\usepackage{float}
\usepackage{todonotes}

\DeclareSymbolFont{cmletters}{OML}{cmm}{m}{it}
\DeclareMathSymbol{v}{\mathalpha}{cmletters}{"76}

\newcommand{\msun}{\,{\rm M_{\odot}}}

\newcommand{\s}{\,{\rm s}}	
\newcommand{\erg}{\,{\rm erg}}

\newcommand{\tmad}{\,{t_{\rm MAD}}}

\newcommand*\diff{\mathop{}\!\mathrm{d}}
\newcommand{\deriv}[2]{\frac{\diff #1}{\diff{#2}}}

\newcommand{\appropto}{\mathrel{\vcenter{
			\offinterlineskip\halign{\hfil$##$\cr
				\propto\cr\noalign{\kern2pt}\sim\cr\noalign{\kern-2pt}}}}}
\usepackage{xcolor}

\shorttitle{Collapsar GRBs Grind their BHs to a Halt}
\shortauthors{Jacquemin-Ide, Gottlieb, Lowell \& Tchekhovskoy}

\begin{document}
	\title{Collapsar Gamma-ray Bursts Grind their Black Hole Spins to a Halt}

	\author[0000-0003-2982-0005]{Jonatan Jacquemin-Ide}
	\email{jonatan.jacqueminide@northwestern.edu}
	\affiliation{Center for Interdisciplinary Exploration \& Research in Astrophysics (CIERA), Physics \& Astronomy, Northwestern University, Evanston, IL 60202, USA}
	
	\author[0000-0003-3115-2456]{Ore Gottlieb}
	\affiliation{Center for Interdisciplinary Exploration \& Research in Astrophysics (CIERA), Physics \& Astronomy, Northwestern University, Evanston, IL 60202, USA}

	\author[0000-0002-2875-4934]{Beverly Lowell}
	\affiliation{Center for Interdisciplinary Exploration \& Research in Astrophysics (CIERA), Physics \& Astronomy, Northwestern University, Evanston, IL 60202, USA}

	\author[0000-0002-9182-2047]{Alexander Tchekhovskoy}
	\affiliation{Center for Interdisciplinary Exploration \& Research in Astrophysics (CIERA), Physics \& Astronomy, Northwestern University, Evanston, IL 60202, USA}
	
\begin{abstract}

    The spin of a newly formed black hole (BH) at the center of a massive star evolves from its natal value due to two competing processes: accretion of gas angular momentum that increases the spin, and extraction of BH angular momentum by outflows that decreases the spin. Ultimately, the final, equilibrium spin is set by the balance between both processes. In order for the BH to launch relativistic jets and power a $ \gamma $-ray burst (GRB), the BH magnetic field needs to be dynamically important. Thus, we consider the case of a magnetically arrested disk (MAD) driving the spin evolution of the BH. By applying the semi-analytic MAD BH spin evolution model of \citet{Lowell2023} to collapsars, we show that if the BH accretes $ \sim 20\% $ of its initial mass, its dimensionless spin inevitably reaches small values, $ a \lesssim 0.2 $. For such spins, and for mass accretion rates inferred from collapsar simulations, we show that our semi-analytic model reproduces the energetics of typical GRB jets, $L_{\rm jet}\sim10^{50}\,\,{\rm erg\,\s^{-1}}$. We show that our semi-analytic model reproduces the nearly constant power of typical GRB jets. If the MAD onset is delayed, this allows powerful jets at the high end of the GRB luminosity distribution, $L_{\rm jet}\sim10^{52}\,\,{\rm erg\,\s^{-1}}$, but the final spin remains low, $ a \lesssim 0.3 $. These results are consistent with the low spins inferred from gravitational wave detections of binary BH mergers. In a companion paper, \citet{Gottlieb2023}, we use GRB observations to constrain the natal BH spin to be $ a \simeq 0.2 $.

\end{abstract}
	
	\section{Introduction}\label{sec:introduction}
    Black holes (BHs) are the product of a massive star core-collapse at the end of its life \citep[collapsar;][]{Woosley1993}. Before the formation of the BH, the stellar core can undergo an intermediate stage during which it collapses into a proto-neutron star (PNS). The large mass reservoir in the stellar core leads to a high mass accretion rate onto the PNS. Once the PNS accretes mass above $ M_{\rm NS,max} \gtrsim 2.2\msun $ \citep{Margalit2017,Aloy2021,Obergaulinger2022}, it collapses to a BH. Observationally, the least massive observed BHs are $ M_{\rm min} \simeq 2M_{\rm NS,max} $, suggesting the presence of a mass gap between $ M_{\rm NS,max} $ and $ M_{\rm min} $ \citep{Bailyn1998,Ozel2010,Farr2011,Mandel2017}. Such gap implies that after their formation and while the stellar collapse is ongoing, BHs continue to accrete mass that is at least comparable to their natal mass, $ M_{\rm NS,max} $ \citep{Belczynski2012,Kovetz2017}.

    As it accretes gas, the BH gains mass and angular momentum, so that its spin can either increase by accretion, or decrease by generating collimated Poynting-flux dominated outflows (jets) that extract BH rotational energy \citep{Penrose1971}. Numerical simulations of rotating collapsars have shown that in the absence of collimated outflows or jets \citep{Shapiro2002,Shibata2002,Fujibayashi2020,Fujibayashi2022}, or if the jets are powered hydrodynamically, rather than by the rotational energy of the BH \citep{MacFadyen1999,Janiuk2008}, the BH spins up by the end of the explosion process to a dimensionless spin $ a \approx 1 $ \citep[see however,][]{Chan2018}. We are unaware of numerical studies that consider both spin-up by accretion and spin-down by jet launching.

    % The BH spin can also be studied observationally, as the BH system emits radiation while it disrupts stars, accretes gas, merges with other compact objects, and/or launches outflows. 
    Several observational techniques have been used over the years to constrain the spin of BHs via electromagnetic (EM) emission of the BH accretion disk, from x-ray reflection spectroscopy \citep[e.g.,][]{Garcia2014} to thermal continuum fitting \citep[e.g.,][]{Zhang1997,mcclintock_black_2014,zhu_testing_2019}. Although these methods suggest that at least some of the BHs are rapidly spinning, these measurements may depend on the poorly understood accretion physics of BHs \citep[see][for reviews]{Middleton2016,Reynolds2021}. A relatively new and more robust technique to infer the BH spin is through gravitational wave detections of binary BH mergers by LIGO/Virgo/KAGRA (LVK). Such studies consistently indicate that pre-merger BHs are slowly spinning \citep{Farr2017,Tiwari2018,Roulet2019,Abbott2020,Hoy2022}.
    
    % using results a set of super eddinghton thick MAD simulations
    % physically motivated model of BH spindown
    % they are able to constraint the efficiency of spin-down by solving the spin down equations
    % THey clearly showed that spin-dow due to MADs is very efficient and only requieres for the BH to accrete around 50% of its initial mass.
    % this could be significant in GRBs as they are very violent events where BHs can accrete a considerable amount of mass.

    Some of the massive progenitors of BHs are stripped envelope stars \citep[e.g.,][]{GalYam2022}. These stars are associated with the detection of $ \gamma $-ray bursts (GRBs), powered by relativistic jets launched from the BH. The enormous energy of those jets indicates that they are powered electromagnetically \citep[e.g.,][]{Lyutikov2003,Leng2014,Liu2015} via the extraction of BH rotational energy by the magnetic fields threading the BH \citep[BZ;][]{Blandford1977}. Therefore, jetted explosions link the birth of BHs and their spin with the formation of relativistic jets in the stellar core, providing a unique opportunity to study BHs through the observables of GRBs.  In a companion paper, \citet{Gottlieb2023}, we argue that GRB observables favor BHs with low natal spins. Here, we analyze the interplay between the BH spin and the jet to study the BH spin evolution, and the final spin at the end of the stellar collapse.

    A spinning BH exchanges angular momentum with its disk-jet accretion system, which results in both hydrodynamic torques through accretion, and magnetic torques through jet launching acting on the BH. This continues until the BH reaches equilibrium spin. In order for the jets to be launched, the BH needs to possess dynamically-important magnetic fields \citep{Komissarov2009}. BHs in this state are in or near the magnetically-arrested disk (MAD) state \citep{bisnovatyi-kogan_accretion_1974,bisnovatyi-kogan_accretion_1976,narayan_magnetically_2003,Tchekhovskoy2011}. Recently, \cite{Lowell2023} used simulations of \cite{Tchekhovskoy2011,Tchekhovskoy2012} to compute the torques applied by a MAD system to a spinning BH. They constructed a semi-analytic model that could reproduce the behavior of magnetohydrodynamic torques of MADs on BHs. They found that MADs spin down BHs to a relatively low equilibrium spin, $a_{\rm eq}\simeq 0.07$, more efficiently than the spin-up by a standard thin disk \citep{Bardeen1970}. 
    For example, an initially maximally spinning BH of $a_0=1$ can reach the equilibrium spin, $a_{\rm eq}\simeq0.07$, by accreting only $50\%$ of its initial mass during the MAD state. By contrast, for a standard thin disk, a minimally spinning BH, $a_0=0$, needs to accrete $\sim 200\%$ of its initial mass to reach the equilibrium spin, $a=1$.

    In this \emph{Letter}, we build on the model of \citet{Lowell2023} to show that the final spin of collapsar BHs associated with Poynting-flux dominated jets is almost inevitably small. In \S\ref{sec:nutshell} we outline the reasoning of why BHs end up being slowly spinning at the end of stellar collapse. In \S\ref{sec:evolution} we present the semi-analytic model of BH spin evolution. In \S\ref{sec:observations} we compare the model with GRB observables to show that for any reasonable stellar collapse scenario, BHs spin down/up to an equilibrium spin of $ a \approx 0.1 $. We summarize and conclude in \S\ref{sec:summary}.

% \section{Results}
    \section{Low final BH spin in a nutshell}\label{sec:nutshell}
% explain spin-down mechanism. Explain that since both torques are powered by accretion directly or indirectly. The spin down only depends on the amount of accreted mass. The accreted mass and initial spin are the only variables of the system.
The angular momentum exchange between the BH, the disk, and the jets leads to magnetohydrodynamic torques on the BH. On one hand, the angular momentum of the matter falling onto the BH drives an accelerating hydrodynamic torque. On the other, the accreting matter advects magnetic fields onto the BH: the magnetic flux threading the BH powers relativistic jets that apply decelerating magnetic torques on the BH. 

In the MAD state, the jet power and EM torque are linked to the accretion power by the jet efficiency, which in turn depends on the BH spin \citep{Lowell2023}. 
Hence, the torques acting on the BH depend only on the accretion rate and the BH spin \citep{Gottlieb2023}. Consequently, the final BH spin depends solely on the initial BH spin, $a_0$ and the total accreted mass, $m_f$. We normalize the total accreted mass by the BH initial mass, $M_0 \equiv M(t=0)   $, to define
\begin{equation}
    \chi \equiv \frac{m_f}{M_{0}}~,
\label{eq:chi}
\end{equation}
where we consider a mass accretion rate $ \dot{m} $ to compute the total accreted mass,
\begin{equation}
    m_f = \int ^{\infty}_{0} \dot{m} \diff t~.
\end{equation}
We stress that $ \chi $ does not represent the mass growth of the BH, but the total accreted mass on the BH. The increment in BH mass is lower than the accreted mass because some of the accreted energy is deposited into launching BH-powered relativistic jets, thus $\chi\neq\frac{M(t\rightarrow\infty)}{M_{0}} $. 
For a sufficiently high accretion rate, e.g. $\chi\gtrsim\frac{1}{2}$, the final BH spin reaches equilibrium spin, $a_f \simeq a_{\rm eq}=0.07$ \citep{Lowell2023}.
Ultimately, the accreted mass is likely related to the stellar mass $ M_\star $. Thus, we define 
\begin{equation}
    \lambda \equiv \frac{m_f}{M_\star}~.
\end{equation}
One might naively expect most of the stellar envelope to fall onto the BH. However, the powerful jets and disk outflows will unbind a considerable fraction of the stellar envelope.

The percentage of the stellar envelope that is accreted by the BH was estimated by \cite{Gottlieb2022c} by measuring the percentage of mass that remains bound at the end of the simulation. They found that $ \sim 20~\s $ after the collapse, $ \lambda $ reaches an asymptotic value of $ \lambda \approx 0.1 $, e.g. $ 10\%$ of the stellar mass will be accreted onto the BH. For such accretion fraction, a stellar envelope of $20\,\,\msun$ and an initial BH mass of $2.5\,\,\msun$ result in $\chi = 0.8 $, as roughly needed for explaining the mass gap, well above the critical value for reaching equilibrium spin. 

\begin{figure*}
    \begin{center}
    \includegraphics[width=0.8\textwidth]{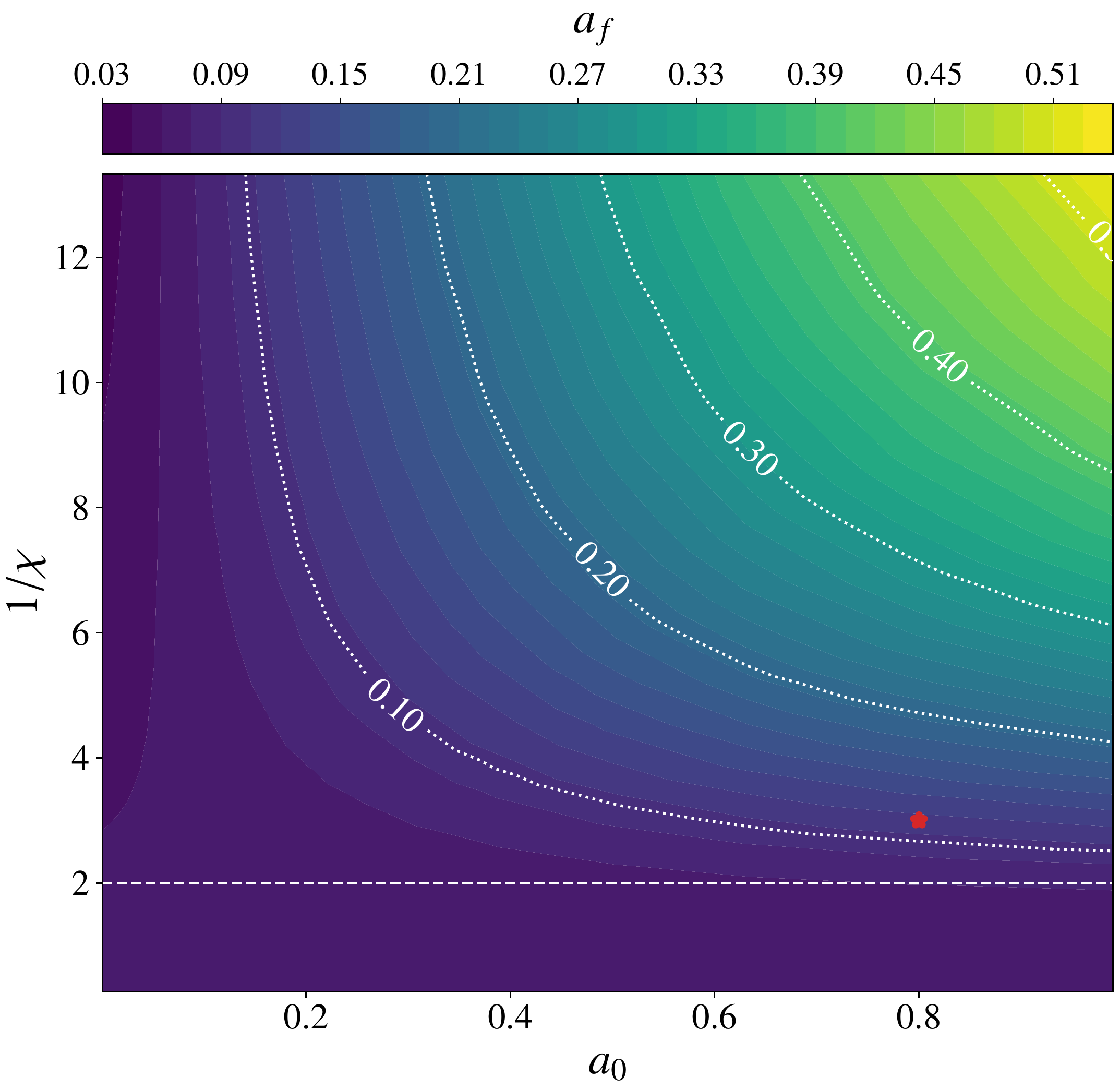}
    \end{center}
    \caption{Final spin of the BH as a function of the inverse of the accreted mass, $1/\chi$ (Eq.~\ref{eq:chi}), and the initial BH spin $ a_0 $. 
    All BHs that accrete more than half of their original mass, i.e. $ 1/\chi < 2 $ showed by the white dashed line, spin down to equilibrium spin, $a_{\rm eq}\simeq0.07$. The simulation of \cite{Gottlieb2022c} with initial spin $a_0=0.8$ and $\chi\simeq0.3$ is marked by the red star. Most BHs spin down to $a\leq 0.3$ as long as they accrete $15\%$ of their initial mass, irrespective of their initial spin. }
    \label{fig:final_spin}
\end{figure*}

To compute the BH spin evolution in time, we model the BH spin by coupling the spin evolution equations to an imposed accretion rate \citep{moderski_black_1996,Lowell2023}

\begin{equation}
    \frac{1}{\dot{m}}\deriv{a}{t} = \frac{s_{\rm MAD}(a)}{M}~,
    \label{eq:dadm}
\end{equation}

\begin{equation}
    \frac{1}{\dot{m}}\deriv{M}{t} = e_{\rm HD}-\eta_{\rm EM}(a)~,
    \label{eq:dlnMdm}
\end{equation}
where the spin-up parameter can be written as
\begin{equation}
    s_{\rm MAD}(a) = \left(l_{\rm HD} - 2a e_{\rm HD} \right) - \eta_{\rm EM}(a) \left(\frac{1}{k(a) \Omega_{\rm H}} - 2a\right)~,
    \label{eq:spinup}
\end{equation} 
where $M$ is the mass of the BH, $\diff m = \dot{m}\diff t$ is the accreted mass, $\eta_{\rm EM}(a)$ is the jet launching efficiency, $e_{\rm HD}$ and $l_{\rm HD}$ are the hydrodynamic energy and angular momentum fluxes, and $ k(a) = \Omega_{\rm F} / \Omega_{\rm H}$ is the angular frequency of the magnetic field lines over the angular frequency of the event horizon.
The numerical values of $e_{\rm HD}$ and $l_{\rm HD}$, and the functions $\eta_{\rm EM}(a)$ and $k(a)$ are taken from \cite{Lowell2023}\footnote{We note that in \cite{Lowell2023} the spin evolution was computed using interpolation between simulation values of $s$, while in our work we use their analytic model. This leads to small differences in the spin evolution.}.

Figure~\ref{fig:final_spin} depicts the final BH spin from multiple spin-down solutions computed with different initial BH spins $ a_0 $, and total accreted mass $ \chi $. All solutions with $1/\chi<2$ reach equilibrium spin, $a_{\rm eq}=0.07$, as was found by \citet{Lowell2023}. The spin-down is also efficient for lower values of accreted mass. For example, even for a small total accreted mass of $ 0.2M_0 $, i.e. $1/\chi=5$, the final BH spin is $ a_f \lesssim 0.2 $. The simulation of \cite{Gottlieb2022c} with an initial spin $a_0=0.8$ and a $\chi\simeq0.3$, marked by the red star in Fig.~\ref{fig:final_spin}, should reach $a_f\simeq0.1$. This demonstrates that even if the natal BH spin is high and $\chi$ is below the critical value of $\chi=0.5$, the spin-down is substantial. We conclude that for any reasonable accreted mass, e.g. $m_f\sim \msun$, collapsar BHs inevitably spin down to low spins of $ a_f \approx 0.1 $, independent of the mass accretion rate.

We stress that the spin evolution model of \cite{Lowell2023} is only valid for an engine that has reached the MAD state and is radiatively inefficient. We verify that the spin evolution model of \cite{Lowell2023} is valid for collapsars by showing that it is compatible with the spin-up parameter in collapsar simulations with $a=0.8$ and $a=0.1$ (see Appendix \ref{B:model_comparison}). In collapsar simulations the system reaches the MAD state relatively fast, $t<1\s$ (see Appendix \ref{B:model_comparison}). However, the system could take longer to reach the MAD state with different initial conditions, as discussed in \S\ref{sec:MAD_delay}.

\section{BH spin evolution}\label{sec:evolution}

The magnitude and time dependence of the accretion onto the BH depend on the stellar mass and density profile, respectively. 1D core-collapse simulations find that density profiles from stellar evolution models, $\rho(r)\propto r^{-2.5}$, flatten prior to the BH formation to $\rho(r) \propto r^{-1.5}$ \citep{halevi_density_2022}. For free-fall of a typical stellar envelope mass, numerical and analytic results show that this power-law leads to a steady BH accretion of $ \dot{m} \gtrsim 10^{-2}\,\, \msun~\s^{-1} $ \citep{Gottlieb2022c,Gottlieb2023}. If we were to extrapolate that rate to typical GRB durations of a few dozens of seconds, the BH would accrete $ m_f \approx \msun $. For $M_0 \simeq M_{\rm NS,max} $, this corresponds to a $50\%$ increase in the BH mass and final spin $ a_f \simeq a_{\rm eq} $.

% \jon{Combine both sections do analytical arguments and then go into model and figure 1. Mention point of Ore simulation in figure 1 discussion}
% \section{Modeling the spin evolution of the BH}\label{sec:model}
 
 % start quickly by mentioning Bevs results again in more detail and mention Figure. so that you can go directly to it. then say that we now detail how we obtain Figure 1 and other results.

 % or make Figure one the first Figure before showing the evolution. could be easier. I don't see the point of showing Figure without anything before.
 % \jon{cite the other paper for equations}

For a roughly constant accretion rate until time $ t_f $, we adopt the following time-dependency of $\dot{m}$,
\begin{equation}
    \dot{m}=\dot{m}_0 \frac{1}{1+e^{t-t_f}},
    \label{eq:mdot}
\end{equation}
where 
\begin{equation}
    t_f = \ln\left(e^{\frac{M_\star\lambda}{\dot{m}_0}}-1\right)~,
    \label{eq:tf}
\end{equation}
is the characteristic accretion duration, the time where the mass reservoir, $m_f = M_\star \lambda$, has been exhausted.
%In our model the total accreted mass during the event is fixed by $\chi=m_f/M_0$, hence $\chi$ is the only parameter that controls the growth and spin-down of the BH. Once $\chi$ is fixed, the accretion rate only controls the duration of the GRB, $t_f$.
For simplicity we adopt a constant $\dot{m}$.
However, we note that a constant accretion rate is not strictly consistent with the simulations of \cite{Gottlieb2023}, which feature a small decrease of $\dot{m}$ with time.  Furthermore, the asymptotic behavior of $\dot{m}$ is not constrained, as discussed in \S\ref{sec:summary}. 

\begin{figure}
	\includegraphics[width=\columnwidth]{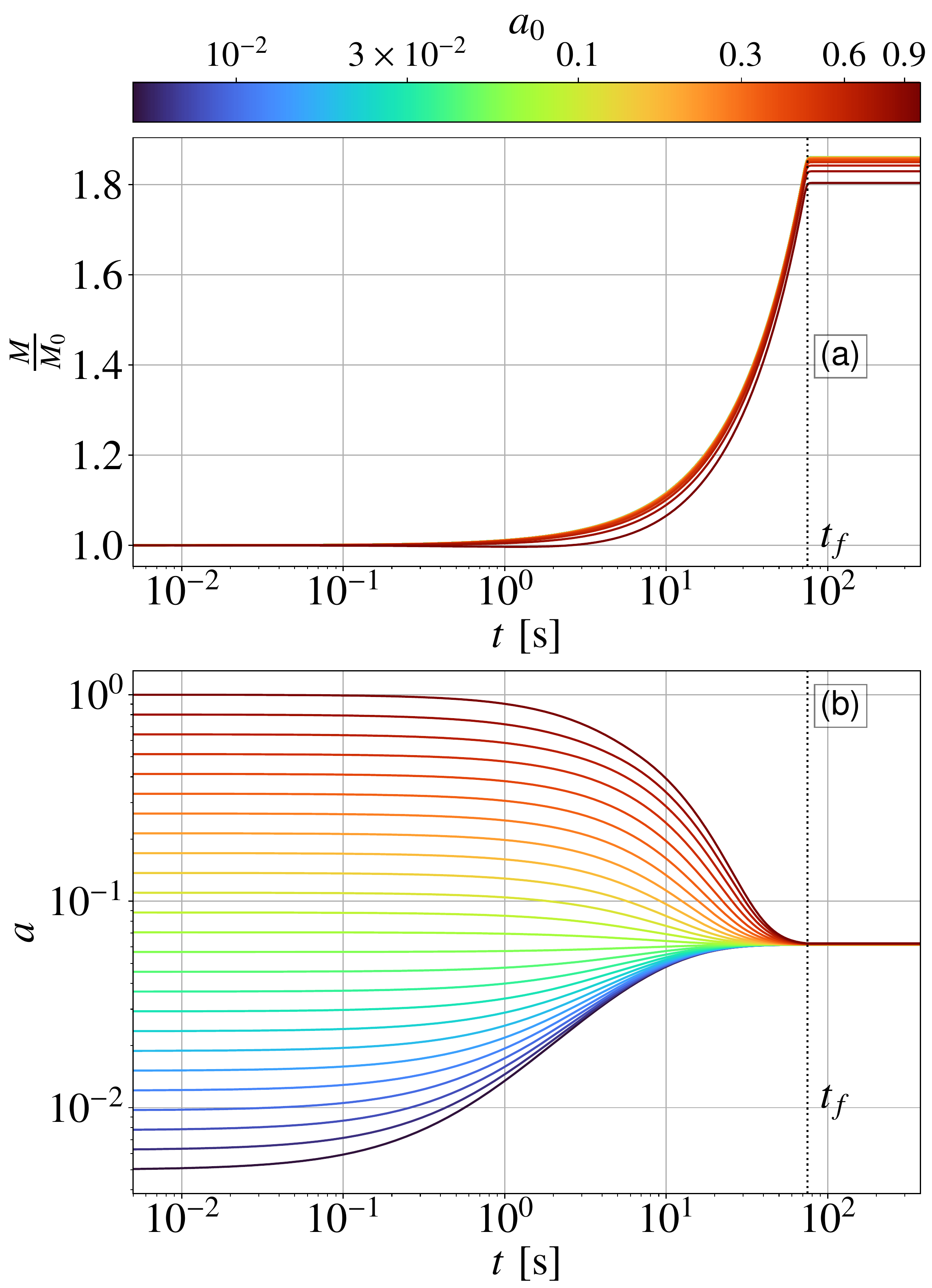}
    \caption{Evolution of $a(t)$ (panel a) and $\frac{M}{M_{0}}(t)$ (panel b) as functions of time and different initial spins $a_0\in[5\times 10^{-3},1.0]$. The time at which accretion stops, $t_f$, is shown by the vertical dotted line.
    All solutions were obtained for $\dot{m}=3\times10^{-2}\,\, \msun\,\s^{-1}$, $\lambda=0.2$, $M_\star = 15 \msun$, and $M_{0}=2.5\msun$, $\chi=1.2$. Smaller spins reach spin equilibrium faster than higher spins.
    }
    \label{fig:diff_spin}
\end{figure}

Figure~\ref{fig:diff_spin} demonstrates the evolution of the BH mass (a) and spin (b) for different initial spins, where $ t = 0 $ is the MAD state activation time, assuming $ \lambda = 0.2 $, $\chi=1.2$. The BH mass saturates once the accretion stops, denoted by $t_f$, whereas the BH spin saturates when $a=a_{\rm eq}$, before $t_f$.
All solutions converge to equilibrium spin, with lower initial spins reaching $ a_f = a_{\rm eq} $ faster. The final BH mass is practically independent of the initial spin. 
%A larger spin will lead to more energy loss due to the jet ($\eta_{\rm EM}$ is larger) and thus a smaller final mass. However, this effect is very small and likely not observable.

% In figure \ref{fig:final_spin} we show the final spin of multiple of spin down solutions computed with different initial spins and total accreted masses. We show the final spin as a function of $1/\chi$ (left axis), the GRB duration (right axis), the initial BH spin. We distinguish that only solutions with $1/\chi<2$ or $\chi>0.5$ reach the equilibrium spin, $a_{\rm eq}=0.07$, as already shown by Beverly et al. . This is consistent with a small $\lambda=m_f/M_\star$. Indeed, a BH with an initial mass of $2.5 M_\odot$ would only have to accrete $9\%$ of the mass of a $15M_\odot$ star to reach spin equilibrium. Furthermore, even for smaller values of $\chi$, $1/\chi=8$, we observe that the maximal spin obtained only reaches $\sim 0.3$. BH spin-down efficiently lower the initial spin.

% Figure \ref{fig:final_spin} is completely independent of the value of $\dot{m}$. The accretion rate will only modify the duration of the GRB, $t_f$. We show that for our chose accretion rate ($\dot{m}=3\times 10^{-2}$) only the solutions with high $\chi$ are consistent with typical GRB durations. For a smaller $\dot{m}$ more GRBs would have a larger duration.

\section{Observables of the spin-down model}\label{sec:observations}
We examine the compatibility of our semi-analytic solution with the observed GRB characteristics: duration $ t_f $, average luminosity $ \langle L_{\rm jet}\rangle $ (Eq.~\ref{eq:AveLjet}), and time evolution $ R_{L_{\rm jet}}$ (Eq.~\ref{eq:Rjet}).
% \jon{Add some paragraph when you say that in this section we explore jet power for strongest jets. COnsider removing $T_{90}$ plots it seems to be useles now}
The jet power is dictated by the mass accretion rate $ \dot{m} $ and jet launching efficiency $\eta_{\rm EM}$, which in turn depends solely on the BH spin \citep{Lowell2023} 
\begin{equation}\label{eq:L}
    L_{\rm jet} = \eta_{\rm EM}(a)\dot{m}c^2~.
\end{equation}
To compute the GRB characteristics, we model the jet propagation inside the star with the semi-analytic model of \citet{Harrison2018}, which relies on the formalism of hydrodynamic jets \citep[][see caveats in \S\ref{sec:summary}]{Bromberg2011}. We adopt the following mass density profile in the star
 \begin{equation}
     \rho \equiv \rho_0 \left( \frac{r}{r_H}\right)^{-1.5}\left(1 - \frac{r}{R_\star}\right)^{3},
     \label{eq:rho_prof}
 \end{equation}
 where $r_H$ is the radius of the BH and $R_\star$ is the radius of the star. The density $\rho_0$ is normalized using the total mass of the star, $M_\star$. Using the jet power from our engine evolution model and the above stellar profile, we compute the jet propagation within the stellar envelope and the breakout time $ t_b $. The jet power at the jet head is computed using the retarded time, $t - z_h/c$, where $z_h$ is the position of the jet head.
 The cumulative energy carried by the jet is approximated to be the jet energy that does not cross the reverse shock before breakout
 \begin{equation}\label{eq:Ej}
     E_{\rm jet}(t) = \int ^{t}_{t_b-R_\star/c} L_{\rm jet} \diff t.
 \end{equation}
We then define $t_1$ and $t_2$ which are the times when $5\%$ and $95\%$ of the jet energy has been released. Thus, $90\%$ of the jet energy is released between in a time scale $T_{90}=t_2-t_1$. 
% \jon{Consider adding curve with $\frac{v}{c}$ profiles}
To represent the characteristic jet power of every solution, we define the average jet power
\begin{equation}\label{eq:AveLjet}
 \langle L_{\rm jet}\rangle \equiv \frac{1}{T_{90}}\int_{t_1}^{t_2}L_{\rm jet}\diff t~.
\end{equation}
To quantify the evolution of the jet power, we define the ratio
 \begin{equation}\label{eq:Rjet}
     R_{L_{\rm jet}} =\left| \frac{\max\left(L_{\rm jet}(t_1\leq t\leq t_2)\right)}{\min\left(  L_{\rm jet}(t_1\leq t\leq t_2)\right)}\right|~.
 \end{equation}
We impose $R_{L_{\rm jet}}\leq 2.5$ so that the jet power remains roughly constant during the GRBs duty cycle \citep[e.g.,][]{mcbreen_cumulative_2002}.

\subsection{BH spin evolution of typical GRBs}\label{sec:common}

 To obtain the characteristic GRB jet power, one needs to consider the highly uncertain $ \gamma $-ray radiative efficiency $ \epsilon_\gamma $. We choose a fiducial value of $ \epsilon_\gamma =0.5 $, so the jet power and energy are $ L_{\rm jet} = L_{\rm jet,obs}/\epsilon_\gamma $, and $ E_{\rm jet} = E_{\rm jet,obs}/\epsilon_\gamma $. The typical range of the GRB jet luminosity and energy are $6\times10^{49}\,\, {\rm erg\,\s^{-1}}\leq \langle L_{\rm jet}\rangle \leq 4\times 10^{51} \,\,{\rm erg\,\s^{-1}}$ and $3\times10^{50}\,\,{\rm erg}\leq \langle E_{\rm jet}\rangle \leq 6\times 10^{51}\,\,{\rm erg}$, respectively \citep{goldstein_estimating_2016}.

\begin{figure}
	\includegraphics[width=\columnwidth]{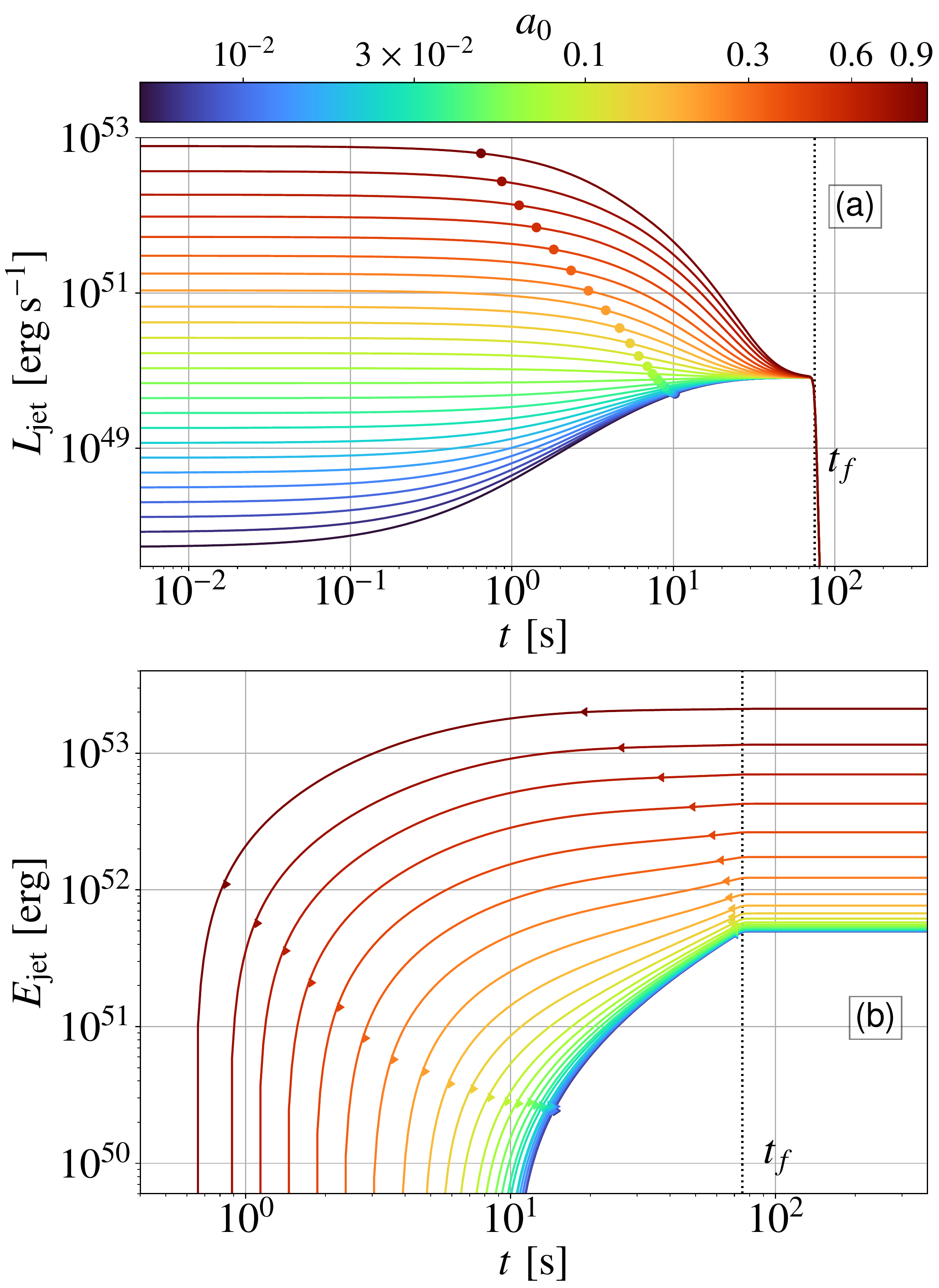}
    \caption{Evolution of $L_{\rm jet}$ (panel a) and $E_{\rm jet}$ (panel b) as functions of time and different initial spins, $a_0\in[5\times 10^{-3},1.0]$. The circles in panel (a) mark $t_b-\frac{R_\star}{c}$, which is the time after which unshocked jet elements can emerge from the star. The left and right triangles in panel (b) represent $t_1$ and $t_2$, respectively.
    The vertical dotted line marks the time at which accretion stops $t_f$. 
    All solutions are computed using $\dot{m}=3\times10^{-2}\,\, \msun\,\s^{-1}$, $\lambda=0.2$, $M_\star = 15 \msun$, $R_\star = R_\odot$, $M_{0}=2.5\msun$, and $\chi=1.2$.}
    \label{fig:diff_spinPow}
\end{figure}

 \begin{figure}
	% To include a Figure from a file named example.*
	\includegraphics[width=\columnwidth]{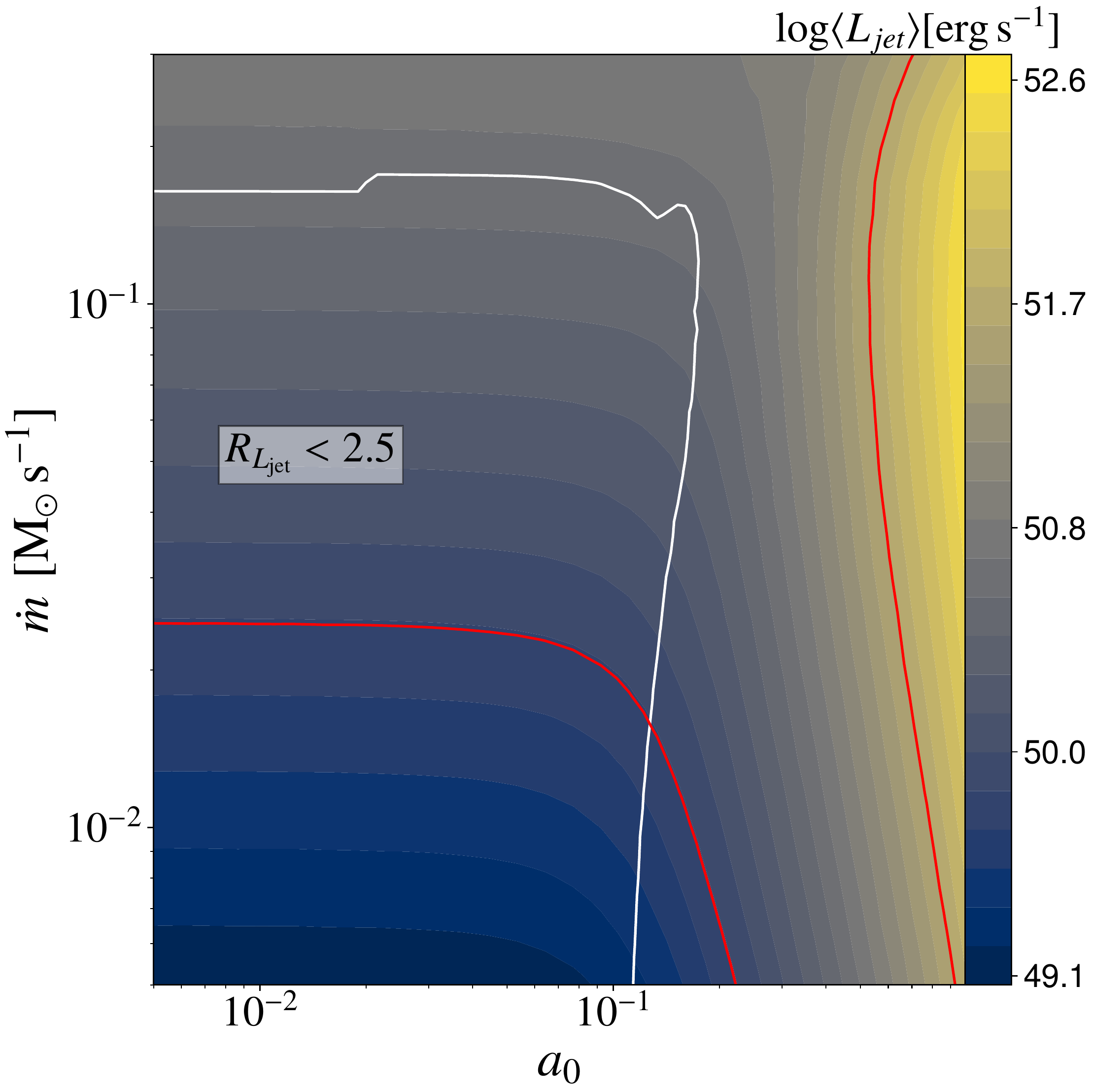}
    \caption{Jet power as a function of $\dot{m}$ and the BH initial spin. The area under the white contour constrains the emerging jets with minimum time evolution in their emission, characterized by $R_{L_{\rm jet}}\leq2.5$. The red lines represent the observational 1$\sigma$ spread around the average jet power \citep{goldstein_estimating_2016}.
    %We choose a fiducial value of radiative efficiency $ \epsilon_\gamma = 0.5 $, by which we divide the observed jet luminosity to obtain an average jet power $\langle L_{\rm jet,obs}\rangle/\epsilon_\gamma = 5\times 10^{50}\,\, \rm erg~s^{-1} $, and a log-normal $1\sigma$ deviation of $\sigma_{L_{\rm jet,obs}} = 0.91 $.
    Only solutions with weak initial spins, $a_0\leq0.1$, and accretion rate of $2\times10^{-2}\,\, \msun\,\s^{-1} \gtrsim \dot{m}\gtrsim 1.5\times 10^{-1}\,\, \msun\,\s^{-1}$  are consistent with both constraints.
    }
    \label{fig:jet_explor}
\end{figure}

Figure~(\ref{fig:diff_spinPow}(a)) delineates the evolution of the jet power and energy for different initial BH spins. We assume $ \dot{m} = 3\times 10^{-2}\msun~\s^{-1} $, which is consistent with our choice of stellar profile and the values measured by \cite{Gottlieb2023}. While higher initial spins lead to higher power and more energetic jets, all BHs reach an equilibrium spin within the typical long GRB duration. Consequently, all jets also converge to the same value, $L_{\rm jet}=\eta_{\rm EM}(a=a_{\rm eq})\times\dot{m}c^2\simeq 8\times10^{49}\,\,\,{\rm erg\,\s^{-1}}$, consistent with the typical GRB jet power.
The unshocked jet element breakout time from the star, $t_b-\frac{R_\star}{c}$, marked by a filled circle, represents the time from which the time evolution in the jet power can be observed. If the BH spin is still evolving considerably at $t>t_b$, there will be visible variations in the observed jet power, in tension with observations.
Fig.~\ref{fig:diff_spinPow}(b) shows the jet energy, as calculated in Eq.~\eqref{eq:Ej}. The energy of all jets launched from BHs with initial spins of $ a_0 \lesssim 0.1 $ is dominated by the energy released when the BH reaches equilibrium spin, $L_{\rm jet}(a=a_{\rm eq})$, and converges to $ E_{\rm jet} \approx 5\times 10^{51}~\erg $ by the end of the GRB, at $t=t_f$. This jet energy is within $ 1\sigma $ of the observed jet energy distribution \citep{goldstein_estimating_2016}. 
% We also show $t_1$ and $t_2$ in Fig.~\ref{fig:diff_spinPow}b. 

\begin{figure*}
 \centering
	% To include a Figure from a file named example.*
	\includegraphics[width=\columnwidth]{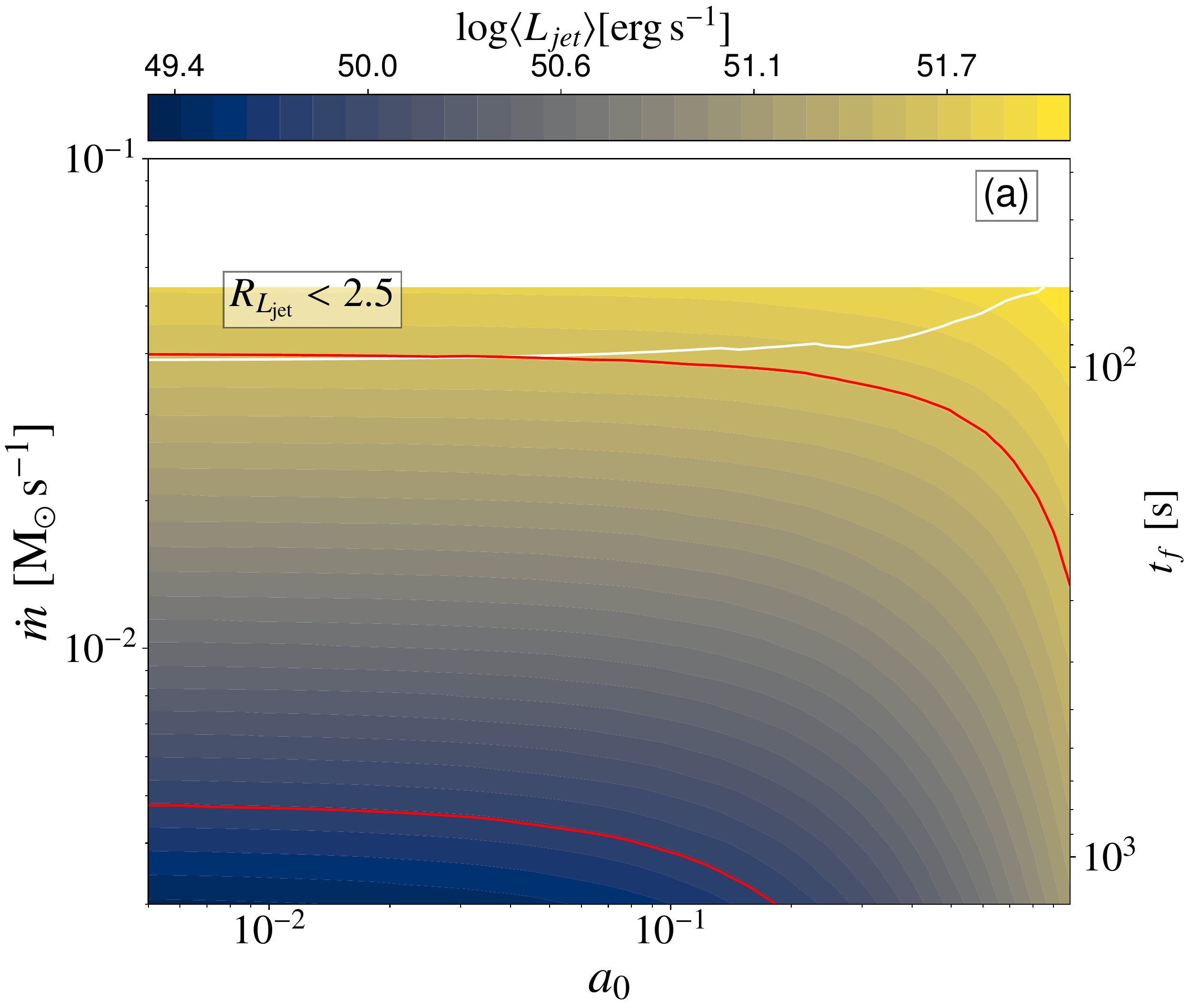}
    \includegraphics[width=\columnwidth]{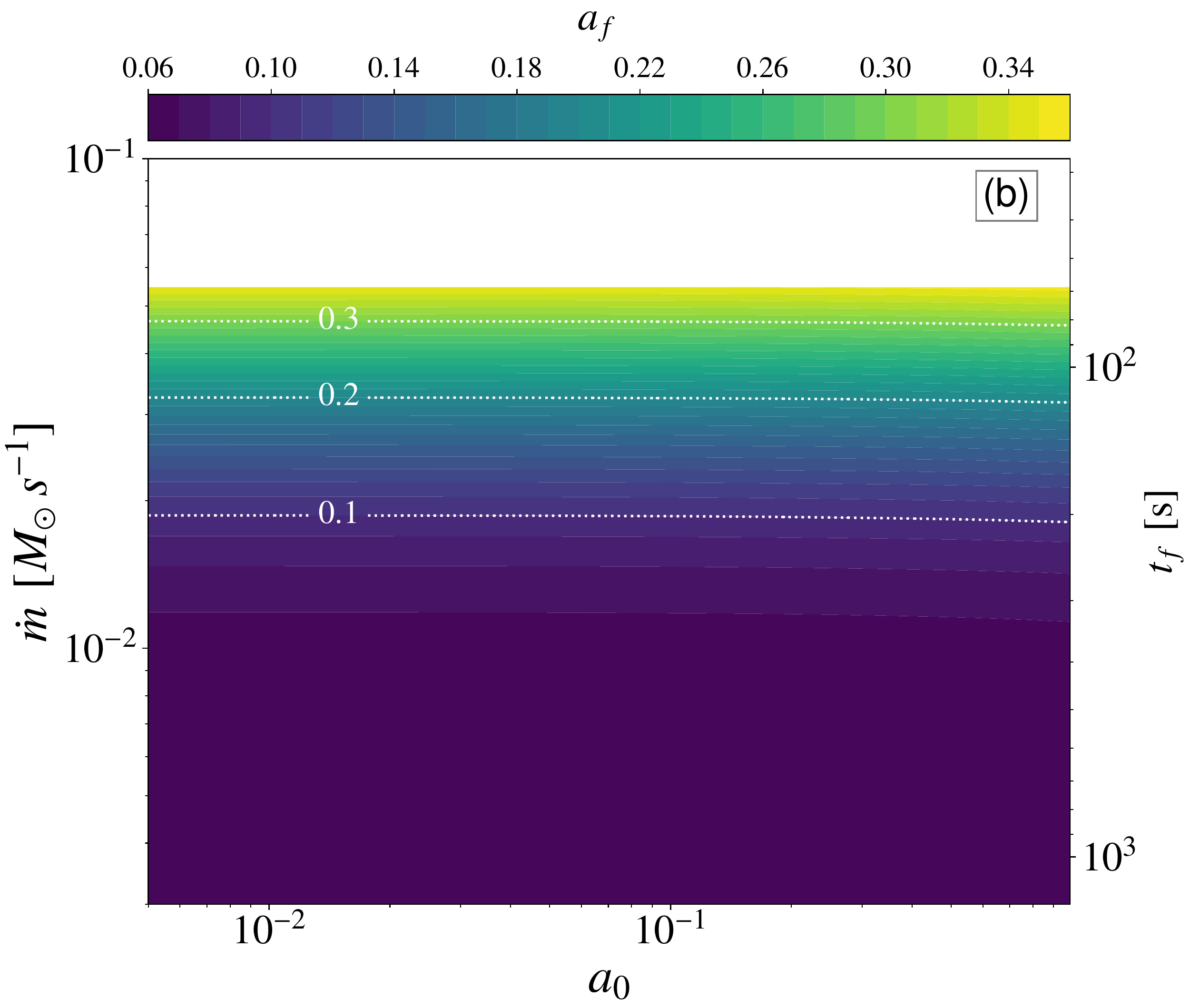}
    \caption{Effect of long $\tmad=70\s$ on the average jet power, $ \langle L_{\rm jet}\rangle$ (panel a) and final spin, $a_f$ (panel b) as functions of the accretion rate $\dot{m}$ and the initial spin, $a_0$. We assume $\chi=1.5$ ($M_\star=25\,\,\msun$, $M_0=2.5\,\,\msun$, $\lambda=0.15$, $R_\star =1\,\,R_{\odot} $). The right axis displays the GRB duration $t_f$, using the accretion rate and $\chi$ from Eq.~\eqref{eq:tf}. We do not plot solutions that have $\tmad>t_f$ (white area). Above the white contour lie the least variable jets, characterized by $R_{L_{\rm jet}}<2.5$. The red lines constrain the observational $ 1\sigma $ spread around the average GRB jet power, assuming $\epsilon_\gamma=0.5$. A large $\tmad$ allows for strong jets, $L_{\rm jet}\geq 5\times 10^{51}\,\,{\rm erg}$, while maintaining small variation, $R_{L_{\rm jet}} $, and small final spins, $a_f\leq0.3$}
    \label{fig:spin_tMAD}
\end{figure*}

Figure~\ref{fig:jet_explor} displays the dependence of the average jet luminosity on the mass accretion rate and initial spin. The average jet luminosity is governed by the more luminous phase of the jet. Thus, for low initial spins, $a_0<a_{\rm eq}$, the average jet power does not depend on the initial spin, and depends only on the accretion rate. This is consistent with Fig.~\ref{fig:diff_spinPow}(a), which shows that low spins quickly reach equilibrium spin, thus $L_{\rm jet} \approx \eta(a_{\rm eq}) \dot{m}c^2 $. This trend is reversed for high initial spins, $a_0>a_{\rm eq}$, where the contour lines are primarily vertical, i.e. the average jet power is dictated by the jet luminosity when the BH spin is $ a_0 $.

Most BH spins and mass accretion rates are consistent within $1\sigma$ with the observed jet power, outlined by the red lines in Fig.~\ref{fig:jet_explor}.
The jet power obtained in the solutions under the white line varies by less than a factor of $2.5$, as shown in Eq.~\eqref{eq:Rjet}. Only solutions with low initial spins, $a_0\lesssim 0.1$, reach equilibrium spin at $ t < t_1 $, thus exhibit a flat jet power curve that could reproduce the observational data (Fig.~\ref{fig:diff_spinPow}a).

For a mass accretion rate of $\dot{m} \gtrsim 10^{-2}\,\, \msun\,\s^{-1}$, a typical GRB jet power is obtained. 
If the mass accretion rate is $\dot{m}<10^{-2}~\msun~\s^{-1} $, weaker jets with $\langle L_{\rm jet,obs}\rangle/\epsilon_\gamma\leq 6\times10^{49}\,\,{\rm erg}$ emerge.

\subsection{BH spin of powerful GRB jets}\label{sec:MAD_delay}

%  \begin{figure}
% 	% To include a Figure from a file named example.*
% 	% Allowable file formats are eps or ps if compiling using latex
% 	% or pdf, png, jpg if compiling using pdflatex
% 	\includegraphics[width=\columnwidth]{Jet_power05.pdf}
%     % \includegraphics[width=0.97\columnwidth]{t_90.pdf}
%     \caption{Jet power as a function of the $1/\chi$ and the initial spin of the BH. We over-plot in white the contours of the least variable jets, characterized by $R_{L_{\rm jet}}<2.5$, they are indicated by the text. We do not plot solutions that have $\tmad>t_f$, that are concentrated in the upper right corner.
%     \label{fig:jet_tMAD}
% \end{figure}
% \jon{Say that high spin not compatible with obs but we still show their final spins in appendix}

Fig.~\ref{fig:jet_explor} shows that powerful GRBs with $L_{\rm jet}/\epsilon_\gamma\geq 5\times 10^{51}\,\,{\rm erg\,\s^{-1}}$ are excluded from the variation constrain. Here we show that the most powerful GRB jets can emerge by delaying the activation of the MAD state (the jet launching).
To investigate the effects of delaying the onset of the MAD state, we introduce the function $\Phi(t)$ that represents the disk state with respect to MAD
\begin{equation}
    \Phi(t) = 1 - e^{-t/\tmad}~,
\end{equation}
where $ \tmad $ is the characteristic time for the disk to become MAD. When $t\ll \tmad$, the disk acts as a standard viscously accreting hydrodynamic disk, and its torques on the BH are modeled with the standard theory of \citet{Bardeen1970}. We write a modified set of spin evolution equations that follow the spin evolution in \citet{Bardeen1970} up until the system becomes MAD
    \begin{equation}
\frac{1}{\dot{m}}\deriv{a}{t} = \frac{s_{\rm MAD}(a)\Phi(t)+s_{\rm Ba}(1-\Phi(t))}{M}~,
\label{eq:dadm_phi}
\end{equation}
and
\begin{equation}
    \frac{1}{\dot{m}}\deriv{M}{t} = e_{\rm Ba}(1-\Phi(t))+\left(e_{\rm HD}-\eta_{\rm EM}(a)\right)\Phi(t)~,
    \label{eq:dlnMdm_phi}
\end{equation}
where $e_{\rm Ba}$ and $s_{\rm Ba}$ are taken from \cite{Bardeen1970}, and $s_{\rm MAD}$ is defined in Eq.~\eqref{eq:spinup}. We reintroduce the jet power in Eq.~\eqref{eq:L} with the magnetic flux saturation parameter as
\begin{equation}
    L_{\rm jet} = \Phi(t)\eta_{EM}(a)\dot{m}c^2.
\end{equation}

When $\Phi(t) = 1$, the magnetic field of the BH and inner disk have saturated and we recover Eqs.~\eqref{eq:dadm} and \eqref{eq:dlnMdm}. When the $t\ll \tmad$, there is no magnetic jet torque breaking the BH. Furthermore, the \citet{Bardeen1970} accelerating hydrodynamic torque on the BH is larger than for a MAD \citep{Lowell2023}. This leads to a far greater equilibrium spin, $a_{\rm eq}=1$. Thus, by having a long $\tmad$, the BH reaches a higher final spin. Solutions with $\tmad\geq t_f$ are excluded from the parameter space, since they would reach the MAD state after the mass reservoir is exhausted. We note that $\Phi(t)$ is a continuous function of $t$, and thus the magnetic flux on the engine and the jet power gradually increases until $\Phi(t)$ saturates. See Appendix \ref{A:More_tmad} for the temporal evolution of solutions with high $\tmad$.

First-principles numerical simulations of collapsars \citep[e.g.,][]{Gottlieb2022a,Gottlieb2022c} show that the disk reaches a MAD state soon after the core collapse. If the disk does not become MAD early on, it energizes an expanding accretion shock that hampers magnetic flux on the disk such that the disk cannot become MAD at later times, disfavoring long $\tmad$. However, those simulations explored only a limited range of magnetic field profiles. It is possible that a low net magnetic flux within the star, or a magnetic flux profile that is concentrated far away from the core, would take a long time to saturate the central engine, leading to a long $\tmad$. Hence, $\tmad$ will depend on the initial magnetic field profile and the magnetic flux transport with the stellar envelope and the disk.
% We note that this scenario is hard to test numerically because low magnetic field strengths are demanding numerically.

Figure~\ref{fig:spin_tMAD} depicts the average jet power (panel a) and final BH spin (panel b) as a function of $\dot{m}$ and $a_0$ for $\tmad=70\s$ and $\chi=1.5$. Fixing $\chi$ in Fig.~\ref{fig:spin_tMAD} leads to $t_f$ being anti-correlated with $\dot{m}$, so high accretion rates lead to fast reservoir depletion. Thus, we exclude the solution with $\dot{m}\gtrsim 5\times 10^{-2}\,\,\msun\,\s^{-1}$ as this entails $t_f\lesssim\tmad$.

In Fig.~\ref{fig:spin_tMAD}(a), the white line delineates $R_{L_{\rm jet}}=2.5$, above which are shown solutions with a low variation, $R_{L_{\rm jet}}<2.5$. The red contours represent the observational $1\sigma$ spread around average jet power \citep{goldstein_estimating_2016}. High jet power, $L_{\rm jet}\geq 5\times 10^{51}\,\,{\rm erg\,\s^{-1}}$, solutions with low variability are obtained above both contours, and are weakly dependent on $ a_0 $. However, very high initial spins, $a_0\gtrsim 0.4$, are excluded since they do not satisfy the variability constraint.

In Figure~\ref{fig:spin_tMAD}(b), the maximum final spin $a_f\simeq 0.35$ is obtained at $t_f\simeq \tmad$, since the MAD state does not have enough time to spin down the BH. At longer $t_f$ (lower $\dot{m}$), the BH can have a substantial spin-down, and reach closer to $ a = a _{\rm eq} $.
Although the final spin is $a_f\lesssim 0.35$ or lower, it peaks at $ a \simeq 0.6 $ before spinning down. This peak is due to the \cite{Bardeen1970} accelerating torque acting on the BH before $\Phi(t)$ saturates ($t<\tmad$). The peak in spin leads to a peak in jet power, so the jet reaches high energies while maintaining a low variation (see Appendix~\ref{A:More_tmad}).
We conclude that delaying the jet activation to $\sim 70\s$ yields jets that are compatible with the most energetic GRBs.

\section{Conclusions and discussion}\label{sec:summary}
\subsection{Summary}
In this \emph{Letter} we show that BH spin evolution to low spins is unavoidable in magnetically arrested collapsars. The final BH spin only weakly depends on the initial spin: it is primarily sensitive to the ratio between the total accreted mass and the BH mass at the onset of the MAD state.
For physically motivated values of accreted mass, this results in a low BH spin. Achieving a high final spin is challenging even for conservative values of BH accretion of $ 20\% $ its initial mass, for which the final spin reaches $a_f \lesssim 0.1$. This will lead to a statistical final spin distribution centered around the equilibrium spin, $a_{\rm eq}=0.07$. 
This is consistent with Bayesian estimation of the BH spin distribution from gravitational wave measurements by LVK \citep{abbott_binary_2019,garcia-bellido_bayesian_2021,edelman_cover_2022}, which constrains the spin distribution of merging BHs to be centered around $a\simeq0.15$. Furthermore, the spin distributions show that highly spinning BHs, $a>0.7$, at least those that end up in merging binaries, should be rare or nonexistent. 
However, it is unclear if BH growth through consecutive mergers or different formation channels would lead to a similar spin distribution.

\begin{table*}
\centering
\begin{tabular}{|l|c|c|c|}
\hline
                  & Weak GRBs                                                             & Most GRBs                                                                                   & Strong GRBs                                                          \\ \hline
Average jet power & $\langle L_{\rm jet,obs}\rangle/\epsilon_\gamma\lesssim 6\times10^{49}\,\,{\rm erg\,\s^{-1}}$ & $6\times10^{49}\,\,{\rm erg\,\s^{-1}}\lesssim \langle L_{\rm jet,obs}\rangle/\epsilon_\gamma \lesssim 4\times 10^{51}\,\,{\rm erg\,\s^{-1}}$ & $\langle L_{\rm jet,obs}\rangle/\epsilon_\gamma\gtrsim 5\times 10^{51}\,\,{\rm erg\,\s^{-1}}$ \\ \hline
Accretion rate    &     $\dot{m}\lesssim 2\times10^{-2}\,\, \msun\,\s^{-1}$                                                                  &          $2\times10^{-2}\,\, \msun\,\s^{-1} \lesssim \dot{m}\lesssim 1.5\times 10^{-1}\,\, \msun\,\s^{-1}$                                                                                    &      $\dot{m}\approx 3\times10^{-2}\,\, \msun\,\s^{-1}$                                                                \\ \hline
Accreted mass     &        $\chi\gtrsim 0.3$                                                               &      $\chi\gtrsim0.3$                                                                                       &           $\chi\gtrsim1.0$                                                           \\\hline
MAD activation    &    $\tmad\ll t_f$                                                                   &           $\tmad\ll t_f$                                                                                  &           $\tmad\sim t_f$, $\tmad< t_f$  \\ \hline   
Initial spin    &    $a_0\lesssim0.1$                                                                   &           $a_0\lesssim0.1$                                                                                  &           $a_0\lesssim0.4$ \\ \hline           
\end{tabular}
\caption{Required parameters for producing a variety of GRB luminosities. The values of jet power are taken from \cite{goldstein_estimating_2016}, and we assume a radiative efficiency $\epsilon_\gamma=0.5$. }
\label{tab:results}
\end{table*}

For consistency, we also check if our BH-powered jets are representative of typical GRBs and do not display any trends in their time evolution.
In Table~\ref{tab:results}, we summarize the required mass accretion and MAD activation times for reproducing the variety of GRB powers. We find that most types of GRBs favor low initial spins, $a_0\leq 0.1$, thanks to their low variation, with the exception of strong jets,$\langle L_{\rm jet,obs}\rangle/\epsilon_\gamma\gtrsim 5\times 10^{51}\,\,{\rm erg\,\s^{-1}}$, that can display little variation up to a higher initial spin, $a_0\simeq0.35$.

In a companion paper \citep{Gottlieb2023}, we show that slowly spinning BH, with $ a_0 \simeq 0.2 $, generate jets with typical GRB powers. GRBs with higher initial spins would exhibit too much variation. Weak GRBs can be explained by $\dot{m}\lesssim 2\times10^{-2}\,\, \msun\,\s^{-1}$, at a low initial spin, $a_0<0.1$ (see Fig.~\ref{fig:jet_explor}).
In order to obtain low variation for powerful GRB jets, we introduce a delayed jet activation time, $\tmad\sim t_f$. This allows the BH to spin up during the first stage of accretion in the absence of jet activity. The rapidly spinning BH then powers a strong jet, and ultimately spins down to $ a_f \lesssim 0.3 $. A long $\tmad$ could be the consequence of a star with a weak magnetic field or a magnetic flux that has a steep radial profile, see \S\ref{sec:MAD_delay}.

\subsection{Limitations}
We discuss the limitations of our semi-analytic model, and how to improve those in future work.

\begin{itemize}

%Jet propagation, and jet evolution
\item In \S\ref{sec:observations} we use a semi-analytic hydrodynamic model to solve the jet propagation in the star. On the one hand, \citet{Gottlieb2020b,Gottlieb2022f} showed that weakly magnetized jets do not develop local hydrodynamical instabilities, and they propagate $ \sim 2 $ times faster than their hydrodynamic counterparts. On the other hand, we consider strongly magnetized jets, which are subject to current-driven instabilities that may slow the jets down. In the absence of numerical modeling of the propagation of such jets, we adopted the numerical solution for hydrodynamic jets by \citet{Harrison2018}. Nevertheless, in the companion paper \citet{Gottlieb2023}, we find that first-principles, strongly magnetized, jets with a typical GRB power fail to retain their relativistic nature upon breakout, due to strong interactions with the star. This is in contrast to hydrodynamic and weakly magnetized models, which do not consider the tilt of the jet launching and kink instabilities. It is thus of utmost importance to generalize jet propagation models based on first-principles simulations.

%MAD termination
\item We delay the activation of the jet by introducing the timescale by which the formation of the MAD is delayed post core-collapse, $\tmad$. However, we did not consider the engine deactivation time, which we set at $t_f $, i.e. when the mass reservoir is exhausted.
Magnetic field transport is responsible for the emergence or end of the MAD state \citep{Tchekhovskoy2011}. Once the magnetic flux saturates on the BH, the inner disk reaches the MAD state. The engine deactivation time then depends on the magnetic field structure of the star and the magnetic field transport after and during the stellar collapse. \cite{jacquemin-ide_magnetic_2021} measured how the magnetic field transport depends on the strength of the initial magnetic field in accretion disks. However, how this results can be generalized to collapsars or to longer timescales is unclear, as the magnetic field transport mechanism remains poorly understood.
Thus, the jet could turn off before $t_f$ if the magnetic reservoir is exhausted or the magnetic structure reorganizes before $t_f$ \citep{Gottlieb2022a}. Long duration global collapsar simulations are needed to better constrain the engine evolution on long timescales.

%Neutrino cooling
\item Our model is built on 3D GRMHD simulations in which the disk is not cooled: they are meant to represent highly super-Eddington accretion systems \citep{Lowell2023}. This is also the case for the simulations of \citet{Gottlieb2022c,Gottlieb2023}. In reality, for the high accretion regimes of GRBs, neutrinos take on as the dominant cooling agent \citep[e.g.,][]{chevalier_neutron_1989,batta_cooling-induced_2014,siegel_collapsars_2019}. Cooling reduces the disk geometric thickness, which may modify the magnetic and hydrodynamic spin-down torques that in turn could shift the equilibrium spin and the timescale on which the BH reaches that spin. \cite{Lowell2023} argued that thin accretion could lead to higher equilibrium spins, $a_{\rm eq}\sim0.3$. Even though this value is larger than the one we consider by a factor of $\sim 3$, it still is a relatively low spin. The biggest uncertainty that thin MADs introduce is the change in the efficiency of the spin-down torques. It is uncertain if a thin MAD requires a higher accreted mass to reach the equilibrium spin. Finally, \cite{Gottlieb2023} find an accretion rate that is not violent enough for neutrino cooling to be very strong for the mass accretion rates involved, so neutrino cooling might not considerably change our results.

%Accretion rate, nonlinear feedback and dependence
\item The simulations of \citet{Gottlieb2023} show that the accretion rate, although slowly varying, is not entirely constant with time. Numerically, due to the limited simulation runtime it is very hard to constrain the accretion rate on very long timescales. Furthermore, it is unclear how the spin-down might affect the non-linear feedback of jets on the accretion rate \citep{Gottlieb2022a}.
Because of this, we do not include any temporal variation in the mass accretion rate in our model. In fact, when we include a slowly varying $\dot{m}$, we find that this makes it hard to satisfy the lack of variation in jet power suggested by observations \citep{mcbreen_cumulative_2002}.
% The accretion rate might show less variations in time for less steep initial stellar density profiles, $\rho\propto r^{-1.35}$ \citep{halevi_density_2022}.

\item First-principles collapsar simulations feature wobbly jets \citep{Gottlieb2022c}, owing to the spontaneous tilt of the accretion disk. Such a wobble would alter the inferred jet energy from the one used in this paper. For example, if the tilt jet angle is $ 0.2~{\rm rad} $ and its opening angle is $ 0.1~{\rm rad} $, then only $ 10\% $ of the jet energy is observed for a given line of sight \citep{Gottlieb2022e}. Thus, the presence of the wobble increases the total GRB jet energy by about an order of magnitude. Such an increase would favor solutions on the border of the parameter space shown in Fig.~\ref{fig:jet_explor}, $\dot{m}\gtrsim 10^{-1} \,\,\msun~\s^{-1}$, and long $\tmad$ values.

\end{itemize}

	\begin{acknowledgements}
    We thank Enrico Ramirez-Ruiz for his thoughtful comments.
    JJ and AT acknowledge support by the NSF AST-2009884 and NASA 80NSSC21K1746 grants.
    OG is supported by a CIERA Postdoctoral Fellowship.
    OG and AT acknowledge support by Fermi Cycle 14 Guest Investigator program 80NSSC22K0031.
    BL acknowledges support by a National Science Foundation Graduate Research Fellowship under Grant No. DGE-2234667. BL also acknowledges support by a Illinois Space Grant Consortium (ISGC) Graduate Fellowship supported by a National Aeronautics and Space Administration (NASA) grant awarded to the ISGC.
    AT was also supported by NSF grants
    AST-2107839, %new short GRB grant
    AST-1815304, %old short GRB grant
    AST-1911080, %accretion grant
    AST-2206471,
    OAC-2031997, %Frontera travel grant
    and NASA grant 80NSSC18K0565. %neutrino francois
    Support for this work was also provided by the National Aeronautics and Space Administration through Chandra Award Number TM1-22005X issued by the Chandra X-ray Center, which is operated by the Smithsonian Astrophysical Observatory for and on behalf of the National Aeronautics Space Administration under contract NAS8-03060.
    This research used resources of the Oak Ridge Leadership Computing Facility, which is a DOE Office of Science User Facility supported under Contract DE-AC05-00OR22725. An award of computer time was provided by the ASCR Leadership Computing Challenge (ALCC), Innovative and Novel Computational Impact on Theory and Experiment (INCITE), and OLCF Director's Discretionary Allocation  programs under award PHY129. This research used resources of the National Energy Research Scientific Computing Center, a DOE Office of Science User Facility supported by the Office of Science of the U.S. Department of Energy under Contract No. DE-AC02-05CH11231 using NERSC award ALCC-ERCAP0022634.

	\end{acknowledgements}
	
	\section*{Data Availability}
	
	The data underlying this article will be shared upon reasonable request to the corresponding author.
	
	\bibliography{refs} 
	\appendix
	% \section{Effects of MAD delay}

    \section{Comparison with spin-down model}
    \label{B:model_comparison}
We compute the spin-up parameter using our semi-analytic model (Eq.~\ref{eq:spinup}) for two collapsar simulations. One simulation has a BH spin $a=0.8$ from \cite{Gottlieb2022c}, and we conduct an additional identical simulation, but with $a=0.1$. We compare the results with the MAD spin-down model from \cite{Lowell2023}, which was calibrated by the GRMHD simulations initialized with a gaseous torus. We can calculate $s$ using 
\begin{equation}
    s = l - 2ae,
\end{equation}
where $l$ and $e$ are the angular momentum and energy fluxes, respectively, onto the BH.

\begin{figure}
     \includegraphics[width=\columnwidth]{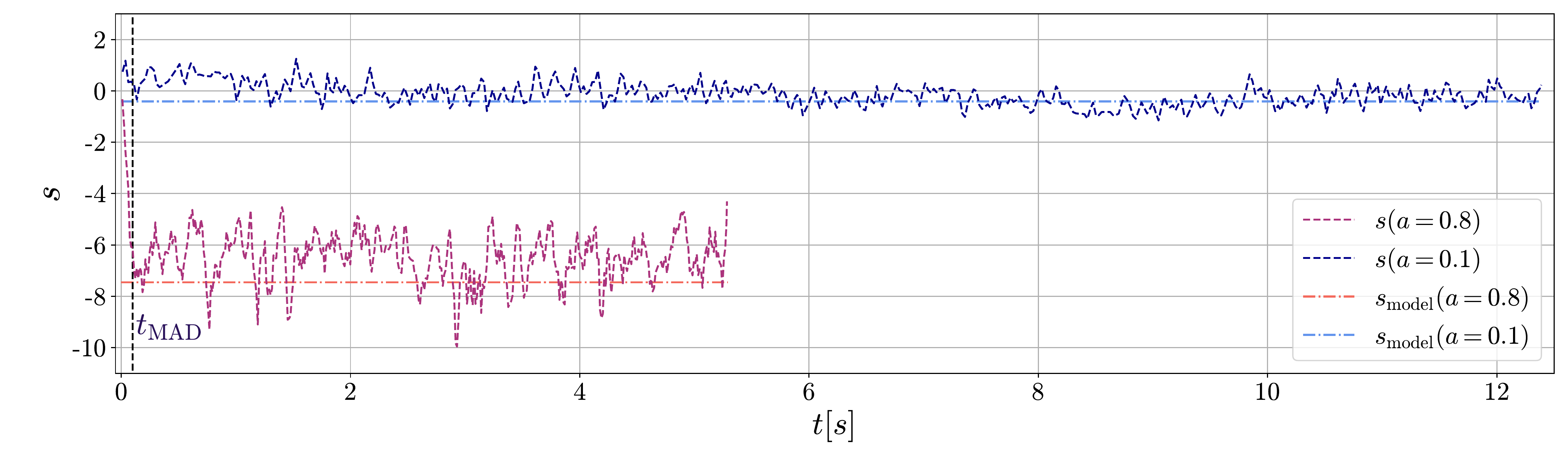}
    \caption{
    Spin-up parameter vs. time for collapsar simulations with BH spin values of $a=0.1$ and $a=0.8$. The pink and dark  blue dashed lines show moving averages of $s$ for $a=0.8$ and $a=0.1$, respectively. The horizontal orange and light blue lines show the values calculated from the MAD model for $a=0.8$ and $a=0.1$, respectively. The time where the system reaches the MAD state in the $a=0.8$ simulation, $\tmad$, is shown by the vertical black line.}
    \label{fig:s_vs_t}
\end{figure}

In Figure~\ref{fig:s_vs_t}, we plot the spin-up parameter vs. time for the two simulations. We calculate the average $s$ for the $a=0.8$ simulation after $t_{\rm MAD}$ and find $s_{\rm avg} \approx -6.5$. For the $a=0.1$ simulation, after $t=1\,\rm{s}$ we find $s_{\rm avg}\approx -0.2$. We show the values of $s$ from the MAD spin-down model, $s_{\rm model}$, with horizontal dashed lines. For $a=0.1$, the model gives $s_{\rm model}=-0.4$, a factor of $\sim 2$ difference, and for $ a = 0.8 $, it gives $s_{\rm model}=-7.5$, a $15\%$ difference. We believe that the weaker spin simulation shows a larger deviation because it is closer to the equilibrium spin, $a_{\rm eq}=0.07$. Hence, the spin-up parameter, $s$, is oscillating close to $0$ and might reach positive values that hinder the average. Furthermore, a factor of $\sim2$ is a small error when compared with the other uncertainties of the model, e.g. accretion rate and magnetic field evolution.
We can conclude that the spin evolution model is consistent with collapsars simulations, showing that the MAD spin evolution model can be used for GRBs. 

% The collapsar spin evolution is consistent with the model, showing that the MAD spin evolution model can be used for GRBs.

    \section{The effects of delaying the onset of MAD state}\label{A:More_tmad}
%     \begin{figure}
% 	% To include a Figure from a file named example.*
% 	% Allowable file formats are eps or ps if compiling using latex
% 	% or pdf, png, jpg if compiling using pdflatex
% 	\includegraphics[width=0.5\columnwidth]{diff_spin_jettMAD.pdf}
%     \caption{Evolution of (top) spin, $a$, and (bottom) $L_{\rm jet}$ as functions of time for different initial spins $a_0\in[5\times 10^{-3},1.0]$. We show $t_1$ and $t_2$ as the two triangles for all different solutions, together they define the $T_{90}$.
%     We also show the time where accretion stops $t_f$ as a black dotted vertical line and the MAD activation time scale $\tmad$ as the blue dashed vertical line. 
%     All solutions were obtained with $\dot{m}=5\times10^{-2}\,\, M_{\odot}\,s^{-1}$, $\lambda=0.2$, $M_\star = 25 M_{\odot}$, $R_\star = R_\odot$, $M_{0}=2.5M_\odot$, $\tmad=60\,\, s$.
%     }
%     \label{fig:jetpow_tmad}
% \end{figure}

\begin{figure}
	% To include a Figure from a file named example.*
	
 \includegraphics[width=0.5\columnwidth]{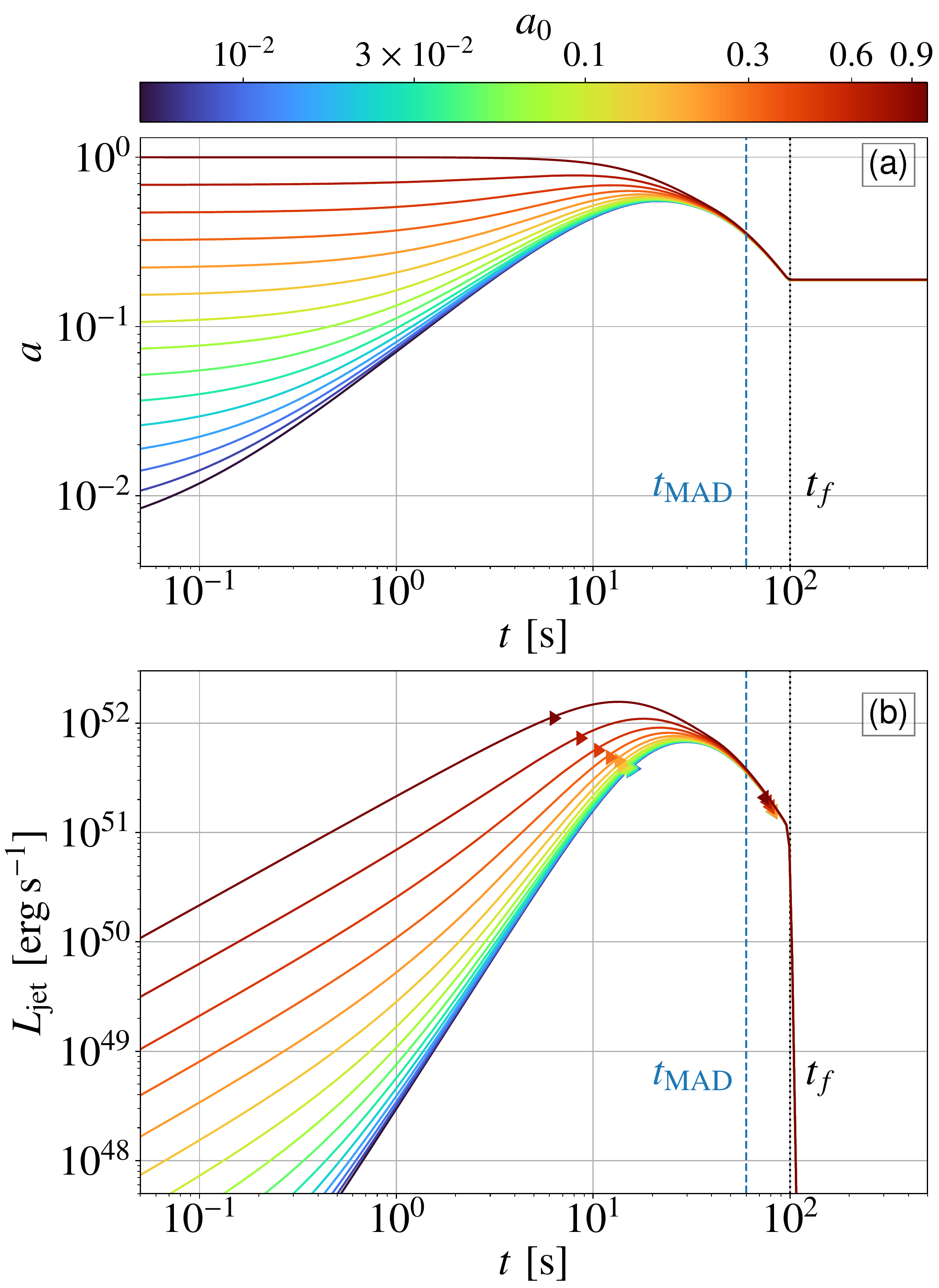}
 \includegraphics[width=0.5\columnwidth]{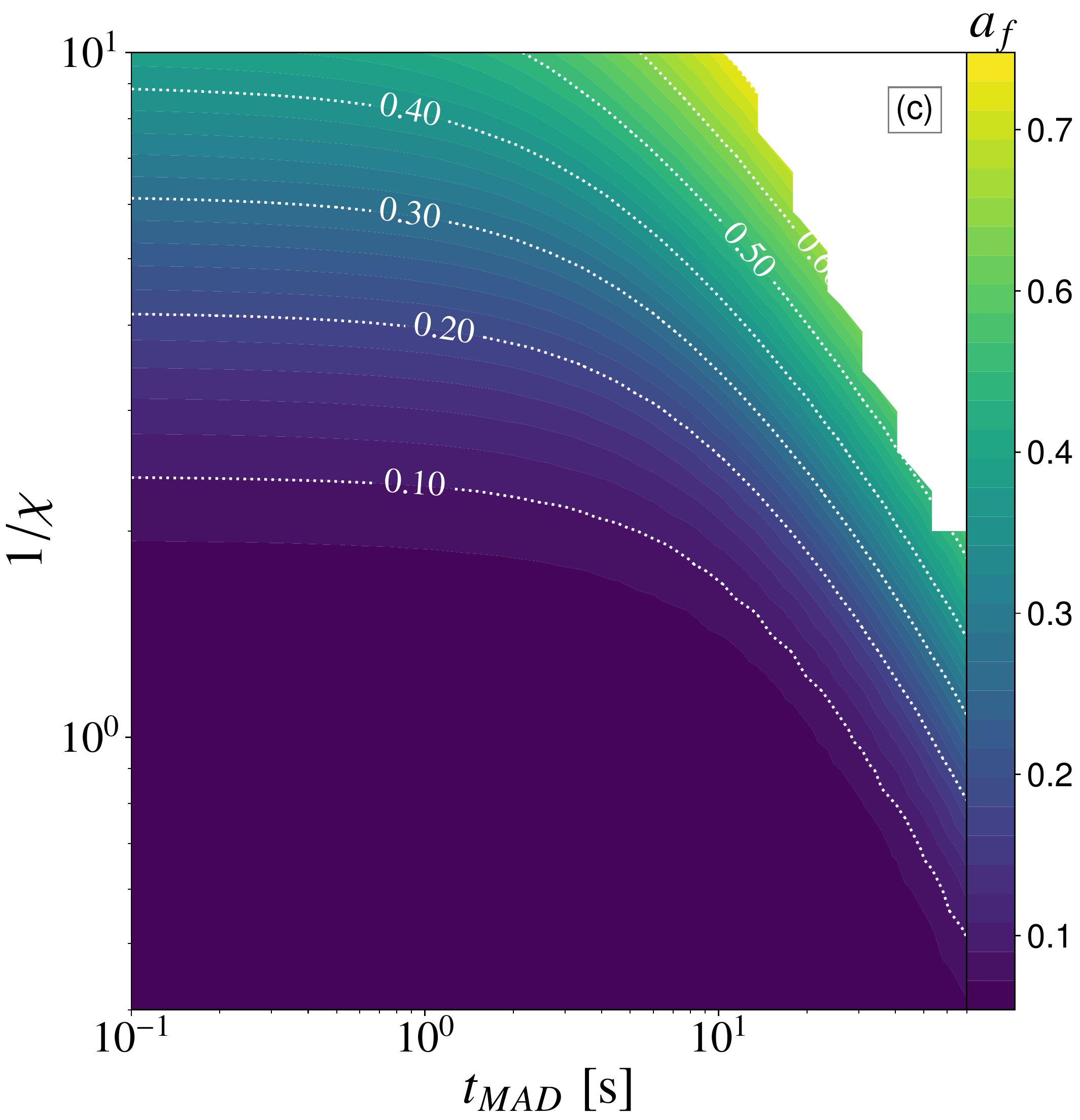}

    \caption{Evolution of  $a$ (panel a) and $L_{\rm jet}$ (panel b) as functions of time and different initial spins $a_0\in[5\times 10^{-3},1.0]$. We also plot $t_f$ (black) and $\tmad$ (blue) as vertical  lines. The left and right triangles in panel (b) represent $t_1$ and $t_2$, respectively.
    (panel c) Final spin, $a_f$, as a function of $1/\chi$ and the MAD activation time scale, $\tmad$ computed for an initial spin $a_0=0.9$. We do not show solutions that have $\tmad>t_f$. For $\chi\geq 0.5$, the MAD activation timescales do not affect the final spin considerably. A high $\tmad$ leads to a higher jet power without considerably modifying the final spin. }
    \label{fig:afinal09}
\end{figure}

Figure~\ref{fig:afinal09}(a,b) shows the jet power and spin profiles for different $ a_0 $ for $\tmad = 60\s $. The left and right triangles in panel (b) show $t_1$ and $t_2$, between which the maximal spin and jet luminosity emerge.
The spin-down torque starts to act well before $\tmad$ because the spin-down torque activates smoothly, mimicking a continuous build up of the magnetic flux on the central BH. The observed jet power is dominated by the times during which the BH spins at $a\sim 0.6$. The solutions still have a low variation, $R_{L_{\rm jet}}<2.5$.

Figure~\ref{fig:afinal09}(c) displays the final spins as a function of $1/\chi$ and $\tmad$ for an initial spin of $a_0=0.9$. In order to achieve high final spins, $a_f\simeq0.6$, $ \chi $ needs to be small and $ \tmad \gtrsim 10\,\s $. Thus, high final spins can be obtained, but requires total accreted mass that is largely inconsistent with the mass gap argument.

% \begin{equation}
%     s(a) = \left(l_{\rm HD} - 2a e_{\rm HD} \right) - \eta_{\rm EM}(a) \left(\frac{1}{k(a) \Omega_{\rm H}} - 2a\right)
%     \label{eq:spinup}
% \end{equation} 
\end{document}